\documentclass[preprint,aps,prx,superscriptaddress,floatfix,longbibliography]{revtex4-2}
\usepackage{amsmath,amssymb,amsthm}
\usepackage{mathtools}
\usepackage{bm}
\usepackage{graphicx}
\usepackage{booktabs}
\usepackage{float}
\usepackage[colorlinks=true,linkcolor=blue,citecolor=blue,urlcolor=blue]{hyperref}
\usepackage[mathlines]{lineno}
\newcommand{\eqdef}{\coloneqq}

\graphicspath{{figures/}{../figures/}}

\begin{document}
\title{Pole-organized cusp catastrophes in locally nonlinear resonator networks}
\author{Kanchan Sarkar}
\affiliation{Institut f\"ur Theoretische Chemie, Universit\"at Ulm, 89081 Ulm, Germany}
\date{\today}

\begin{abstract}
A resonator with one localized cubic nonlinearity can acquire additional cusp
pairs when it is coupled to an otherwise linear network. At first-harmonic
order, the cusp locus is a phase contour of the driving-point receptance
$G(\Omega)$ at the nonlinear coordinate. We resolve that contour into the
contribution of an active pole and its background. For a lightly damped,
isolated modal pole on a slowly varying, nearly real background, cusp-pair
birth occurs once the participation of that pole at the nonlinear coordinate
exceeds a closed-form threshold set by its damping half-width and that
background. On the weak-mixing hardening branch generated by a pole above the
primary resonance, the birth coupling then grows as the square root of the
damping at fixed mode separation, and of the mode separation at fixed damping.
The curvature of the full cusp contour determines whether
the local fold set opens a gap (beaks type) or creates an isolated loop
(lips type). The isolated-pole approximation reduces this classifier to a
background-receptance rule whose range of validity is quantified on 100
random networks. The same ensemble tests the pole ladder numerically, while
multi-harmonic continuation follows the whole reconnection
sequence---connected band, gap, isolated lens, none---in the original
equations at the fold of the birth locus, where the classifier is least
protected. The construction
also extends to coupled-mode Kerr networks, where the driving-point response
sets the intrinsic fold geometry and the drive-port transfer response sets
its accessibility.
\end{abstract}

\maketitle


\section{Introduction}\label{sec:intro}

A periodically driven Duffing resonator can support three period-one
responses over a finite range of frequency and drive. Two saddle-node folds
bound this range and meet at a cusp, where their separation vanishes with
the familiar $3/2$ law
\cite{Thom1972,Arnold1981,GolubitskySchaeffer1985,PostonStewart1978,
HolmesRand1976,ThompsonStewart2002}. With several linear modes, the response
seen by the nonlinear coordinate contains additional resonances and
antiresonances. As those spectral features move with coupling or detuning,
new fold sectors can appear and reconnect.

We consider networks in which only one coordinate carries a homogeneous
cubic force; the rest of the network is linear and otherwise unrestricted in
size or coupling topology. This setting includes coupled mechanical
resonators, photonic molecules with one Kerr-active mode, and microwave
circuits containing a Josephson element. In the last setting, black-box
descriptions already characterize the linear electromagnetic environment by
the response presented to the junction \cite{NiggEtAl2012}. The question here
is whether that local linear response also determines where new fold
catastrophes can occur.

Classical describing-function theory expresses jump-resonance conditions
through the complex frequency response of the linear plant and has long
included disconnected or island-like response structures
\cite{WestDouceLivesley1956,Hatanaka1963,FukumaMatsubara1966,
FukumaMatsubara1978PartI,HiraiSawai1978,KoenigsbergDunn1975}. Modern
describing-function studies have also used this viewpoint to design
multi-jump and multiple-hysteresis responses
\cite{BuscarinoFamosoFortunaFrasca2020,BucoloBuscarinoFortunaFrasca2020}.
Receptance-based methods reduce systems with localized nonlinear elements to
response quantities at the nonlinear coordinates \cite{RenBeards1994}.
Catastrophe and singularity methods classify the cusps and higher
reorganizations of nonlinear forced responses
\cite{HolmesRand1976,Cirillo2017,HabibCirilloKerschen2018,DekemeleHabib2025}, while
effective-equation and reduced-order approaches describe metamorphoses and
detached branches in coupled systems
\cite{KyziolOkninski2011,KyziolOkninski2017,KyziolOkninski2020,
MartinezDuqueVasquezRomero2024,Ponsioen2019,Melot2025}. In the two-mode
geometry used for the examples below, the appearance and disappearance of
detached response curves under parameter change has been analyzed directly
\cite{GattiBrennanKovacic2010,GattiKovacicBrennan2010,GattiBrennan2011,Gatti2016}
and observed experimentally \cite{GattiBrennan2017}; bifurcation tracking in
harmonic balance locates such folds and cusps in the full nonlinear response
\cite{DetrouxEtAl2015}. Both routes build the nonlinear response first. The
question here is how much of that geometry is already fixed by the linear
response, and which pole of the network fixes it. A complementary
strand classifies the stationary-state topology of driven-dissipative
nonlinear systems through an index on their Floquet semiclassical equations
\cite{VillaEtAl2025}. For a single localized cubic, these strands meet in one complex response
function: the driving-point receptance at the nonlinear coordinate.

This paper answers that question pole by pole. Resolving $G$ into an active
pole and its background gives a closed-form threshold: a visible pole nucleates
an additional cusp pair once its participation at the nonlinear coordinate
exceeds a value set by its damping and the surrounding response, with explicit
scaling in damping and in mode separation. The sign of that background fixes
the local beaks- or lips-type geometry, a $100$-network ensemble bounds where
the two reductions hold, multi-harmonic continuation follows the reconnection
sequence through the fold of the birth locus where the classifier is weakest, and the same algebra carries to coupled-mode and
Josephson networks.

The construction those results refine was established in a short
Letter~\cite{Sarkar2026Letter}: at fundamental-harmonic order, eliminating the
nonlinear amplitude places the cusps on a fixed physical phase contour of
$G(\Omega)$, a projection fold of that contour creates a cusp pair, and the
contour curvature gives its local type. The Letter also reported the
harmonic-comb correction and inverse design. Earlier work
\cite{Sarkar2026PartI} concerns a different singular object, the fold of an
exact higher-harmonic zero manifold.

\begin{figure}[!tbp]
  \centering
  \includegraphics[width=\columnwidth]{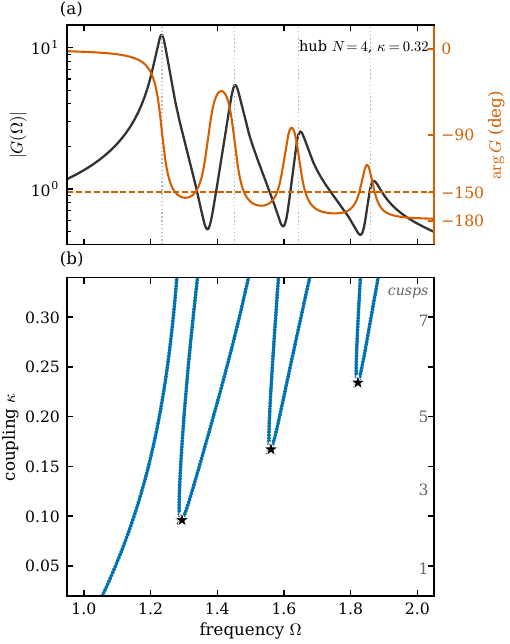}
  \caption{Pole-resolved overview for a four-resonator hub with satellite
frequencies $1.25$, $1.5$, and $1.75$.
(a) The driving-point receptance exposes the visible modal resonances at the
nonlinear coordinate. (b) As coupling increases, successive satellite poles
generate resolved cusp-pair births (stars), so the cusp count rises
$1\to3\to5\to7$. Section~\ref{ssec:kstar} derives the isolated-pole
threshold and Section~\ref{sec:cascade} tests the ladder in larger networks.}
  \label{fig:cascade}
\end{figure}

The theory concerns the period-one fold and cusp set. The reductions below
are exact for the first-harmonic balance equations, not for the full
time-dependent dynamics. The localized cubic form and its sign are assumed
known; the sign selects the physical phase contour. Higher harmonics,
nonlinear dissipation, subharmonic or quasiperiodic responses, and branch
stability require additional calculations. We use multi-harmonic
continuation and Floquet analysis to test those limits.
Figure~\ref{fig:cascade} previews the pole ladder developed below.

\section{First-harmonic reduction to the local receptance}\label{sec:model}
For a localized cubic, the first-harmonic equations reduce to the response
seen at the nonlinear coordinate. We state that reduction first because it
defines the linear quantity used by the pole theory below.

\subsection{General reduction}\label{ssec:receptance}
Let $e_1$ select the only coordinate carrying a cubic force. Write the
first-harmonic response as $x(t)=\operatorname{Re}[\mathbf{X}e^{i\Omega t}]$,
so $\mathbf{X}$ carries physical amplitudes rather than two-sided Fourier
coefficients. For a collocated drive, first-harmonic balance gives
\begin{equation}
Z(\Omega)\,\mathbf{X}
 + \tfrac34\beta_1 |X_1|^2X_1 e_1
 = F e_1,
\qquad
X_1=e_1^T\mathbf{X},
\label{eq:hb_physical}
\end{equation}
where $Z(\Omega)=K-\Omega^2M+i\Omega C$. Solving the linear coordinates and
projecting back onto $e_1$ yields
\begin{equation}
F=X_1\left[\frac{1}{G(\Omega)}+\tfrac34\beta_1|X_1|^2\right],
\qquad
G(\Omega)=e_1^T Z(\Omega)^{-1}e_1 .
\label{eq:receptance}
\end{equation}
This is an exact scalar reduction of the first-harmonic balance equations:
no linear mode has been discarded after the projection onto the fundamental.
The same idea underlies receptance methods for systems with localized
nonlinear elements \cite{RenBeards1994}. Here it will be used to remove the
nonlinear amplitude at a cusp. Once the localized cubic form and its sign are
specified, the first-harmonic cusp locus depends on the linear environment
through $G$ alone.

The scalar closure survives arbitrary linear network size and topology, and
also more general linear time-invariant damping through the corresponding
response function. It does not survive an additional independent nonlinear
coordinate or a cross-cubic term, which introduces further nonlinear
directions (Section~\ref{ssec:open}).

The drive need not be collocated with the nonlinearity. If it acts along a
vector $d$, define the transfer receptance
$H(\Omega)=e_1^T Z(\Omega)^{-1}d$. Equation~\eqref{eq:receptance} becomes
\begin{equation}
F\,\frac{H(\Omega)}{G(\Omega)}
 =X_1\left[\frac{1}{G(\Omega)}+\tfrac34\beta_1|X_1|^2\right].
\label{eq:receptance_drive}
\end{equation}
Thus $G$ fixes the intrinsic fold and cusp condition in frequency and in
the nonlinear-coordinate amplitude, whereas $H$ converts the corresponding
intrinsic forcing into the external drive required at the chosen port. A
transfer zero of $H$ at a cusp frequency leaves the intrinsic organizer
unchanged but makes that cusp inaccessible from that port.
This distinction becomes useful in Section~\ref{ssec:cmt}.

\subsection{Two-mode absorber}\label{ssec:model}
Our detailed mechanical example is the two-degree-of-freedom Duffing absorber
\begin{align}
\ddot{x}_1+2\gamma_1\dot{x}_1+\omega_1^2x_1
 +\kappa(x_1-x_2)+\beta_1x_1^3 &=F\cos\Omega t,
\label{eq:eom1}\\
\ddot{x}_2+2\gamma_2\dot{x}_2+\omega_2^2x_2
 +\kappa(x_2-x_1)&=0,
\label{eq:eom2}
\end{align}
with $(\omega_1,\omega_2)=(1,1.25)$,
$(\gamma_1,\gamma_2)=(0.015,0.02)$, $\beta_1=0.147$, and
$\kappa\in[0.04,0.30]$ unless stated otherwise. Here $\gamma_j$ is a
damping rate (the resonance half-width in the lightly damped limit), not a
dimensionless damping ratio; for a physical mode of quality factor $Q_j$,
$\gamma_j/\omega_j\simeq(2Q_j)^{-1}$. The cubic acts only on the driven
coordinate.

For proportional modal damping, diagonalizing the linear stiffness gives the
explicit two-pole form
\begin{equation}
G(\Omega)=\frac{\phi_{a,1}^2}{D_a(\Omega)}
          +\frac{\phi_{b,1}^2}{D_b(\Omega)},
\qquad
D_m(\Omega)=\omega_m^2-\Omega^2+2i\gamma_m\Omega .
\label{eq:closed_form}
\end{equation}
This representation is useful for the isolated-pole calculation of
Section~\ref{ssec:kstar}. Numerical comparisons with the original equations
use instead the exact physical-coordinate receptance $e_1^T Z^{-1}e_1$.
At the baseline, replacing the modal form by the exact receptance moves the
beaks threshold from $(\Omega^\ast,\kappa^\ast)=(1.30100,0.11837)$ to
$(1.30088,0.11896)$, about $0.5\%$ in $\kappa^\ast$, without changing its
type. The difference reaches about $6\%$ in $\kappa^\ast$ for the
$\omega_2=0.95$ lips example, so closed-form results below are labeled as such.

\subsection{Cusp condition}\label{ssec:cuspform}
Write $1/G=\rho+i\sigma$ and $y=|X_1|^2$. Taking the modulus of
Eq.~\eqref{eq:receptance} gives
\begin{equation}
F^2=y\bigl[(\rho+cy)^2+\sigma^2\bigr],
\qquad c=\tfrac34\beta_1 .
\label{eq:cubic}
\end{equation}
The two saddle-node roots in $y$ exist when $c\rho<0$ and
\begin{equation}
\rho^2>3\sigma^2 .
\label{eq:bistability}
\end{equation}
They coalesce at the cusp boundary $\rho^2=3\sigma^2$, the discriminant
from which the pole threshold below will be derived.

For a passive collocated network with $C\succeq0$, the driving-point
receptance satisfies $\operatorname{Im}G(\Omega)\le0$ for $\Omega>0$.
Together with $c\rho<0$, the cusp boundary selects one physical
phase level,
\begin{equation}
\arg G(\Omega_c)=
\begin{cases}
-150^\circ, & \beta_1>0 \quad (\text{hardening}),\\[2pt]
-30^\circ, & \beta_1<0 \quad (\text{softening}).
\end{cases}
\label{eq:phasecontour}
\end{equation}
Within first-harmonic balance, the magnitude of $\beta_1$ sets the cusp
amplitude and drive scale but not its frequency; its sign selects the
physical contour.

\subsection{Why the second pole cannot be eliminated}\label{ssec:adiabatic}
Far from the upper mode, Eq.~\eqref{eq:closed_form} reduces to a single-pole
Duffing response. That approximation necessarily loses the phenomenon of
interest here. The additional cusp pair is created near the second visible
pole, precisely where its contribution to $G$ is not perturbative. In the
two-mode absorber an adiabatic elimination reproduces the lower cusp closely
but misses the upper pair altogether. We therefore retain the full
receptance, and use the isolated-pole approximation only locally, after the
relevant pole has been kept explicitly.

\subsection{Accuracy of the first-harmonic organizer}\label{ssec:validity}
The projection onto the fundamental is the approximation behind
Eqs.~\eqref{eq:receptance}--\eqref{eq:phasecontour}. Away from a cusp,
$\varepsilon=|\beta_1||X_1|^2/\omega_1^2$ is the usual weak-anharmonicity
measure. At a cusp the double-root condition changes the ordering: the
relative leading feedback through the generated third harmonic is instead
set by the receptance comb,
\begin{equation}
\eta_3=\frac{\sqrt3}{9}
\left|\frac{G(3\Omega)}{G(\Omega)}\right| .
\label{eq:eta3}
\end{equation}
The all-harmonic rank-one closure and the derivation of
Eq.~\eqref{eq:eta3} were given in the Letter \cite{Sarkar2026Letter}. We use
$\eta_3$ here only as a diagnostic for the first-harmonic organizer. At the
baseline cusp it is $1.4\times10^{-3}$; the corresponding corrected frequency
shift is $-2.4\times10^{-5}$, below the uncertainty of the cusp estimator.
Section~\ref{sec:hbval} tests the leading geometry directly against converged
multi-harmonic solutions, including cases where the approximation is less
accurate.

\section{Cusp-pair birth and the isolated-pole threshold}\label{sec:beaks}
For the two-mode absorber, the physical phase contour has a persistent lower
crossing and, above a critical coupling, two additional crossings near the
upper pole. The exact receptance places their birth at
$(\Omega^\ast,\kappa^\ast)=(1.30088,0.11896)$; the modal approximation
\eqref{eq:closed_form} gives $(1.3009974,0.1183654)$ and is used where a
closed form is required. The Letter identified this birth with a projection
fold of the receptance cusp contour \cite{Sarkar2026Letter}. We recall the
local form briefly and then resolve the condition into an active pole and its
background.

\subsection{Projection fold of the cusp locus}\label{ssec:beaksdef}
Define
\begin{equation}
f(\Omega,\kappa)=\rho^2(\Omega,\kappa)-3\sigma^2(\Omega,\kappa),
\qquad \rho+i\sigma=\frac{1}{G(\Omega,\kappa)} .
\label{eq:fdef}
\end{equation}
A generic point of $f=0$ is one cusp. A pair is created when this curve has
a fold under projection onto the varied network parameter,
\begin{equation}
f=0,\qquad \partial_\Omega f=0,
\label{eq:birthcondition}
\end{equation}
with $\partial_\kappa f\neq0$ and $\partial_\Omega^2f\neq0$ for a generic
birth. In the modal two-pole approximation,
Eq.~\eqref{eq:birthcondition} gives
\begin{equation}
(\kappa^\ast,\Omega^\ast)=(0.1183654,1.3009974),
\label{eq:beakspoint}
\end{equation}
while the exact physical-coordinate receptance gives
$(0.11896,1.30088)$. The difference is the modal-damping approximation,
not a change in the organizing condition.

Let $\delta\Omega=\Omega-\Omega^\ast$ and
$\delta\kappa=\kappa-\kappa^\ast$. Near the birth,
\begin{equation}
f\simeq a_\Omega\,\delta\Omega^2+b_\kappa\,\delta\kappa,
\qquad
a_\Omega=\tfrac12\partial_\Omega^2f,
\quad b_\kappa=\partial_\kappa f .
\label{eq:beaksexp}
\end{equation}
For the modal example, $a_\Omega=1370$ and $b_\kappa=-9.00$; the exact
physical-coordinate receptance gives $1315$ and $-8.69$ at the same birth.
Only the sign of $a_\Omega$ classifies. Its magnitude is dimensional---$f$
has the dimensions of $(\text{stiffness})^2$---so the ratios of $a_\Omega$
quoted below compare curvatures within the single nondimensionalization used
throughout, and are not absolute. The two cusp frequencies separate as
\begin{equation}
\Delta\Omega\simeq
2\sqrt{-b_\kappa/a_\Omega}\,\sqrt{\kappa-\kappa^\ast}
=0.162\sqrt{\kappa-\kappa^\ast} .
\label{eq:beaksroot}
\end{equation}
The square-root exponent is the generic projection-fold scaling; the
prefactor depends on receptance derivatives. The Letter illustrates this
geometry explicitly \cite{Sarkar2026Letter}. Here we use it as the starting
point for a pole-resolved birth threshold.

\subsection{Closed-form isolated-pole threshold}\label{ssec:kstar}
The projection-fold condition becomes more informative when the active
resonance is an isolated, lightly damped modal pole. Near the upper mode,
write
\begin{equation}
G(\Omega)\simeq G_a(\omega_b)+
\frac{\phi_{b,1}^2}{D_b(\Omega)},
\qquad
D_b(\Omega)\simeq2\omega_b(-\delta+i\gamma_b),
\label{eq:polebackground}
\end{equation}
with $\delta=\Omega-\omega_b$. The factor $\phi_{b,1}^2$ is the weight of
mode $b$ in the driving-point receptance: for mass-normalized modes it is the
squared component of that mode on the nonlinear coordinate. A dark mode has
$\phi_{b,1}=0$ and cannot reach the threshold below.

Let
\[
A=\frac{\phi_{b,1}^2}{2\omega_b},\qquad
g=\operatorname{Re}G_a(\omega_b),
\]
and neglect $\operatorname{Im}G_a$ at leading order. For the hardening
upper-pole branch used below, $g<0$ and the physical cusp condition is
$\operatorname{Re}G-\sqrt3\,\operatorname{Im}G=0$. Substituting
Eq.~\eqref{eq:polebackground} gives
\begin{equation}
g(\delta^2+\gamma_b^2)-A\delta+\sqrt3\,A\gamma_b=0 .
\label{eq:pole_quadratic}
\end{equation}
The two cusp roots merge when the discriminant vanishes,
\begin{equation}
A^2-4g^2\gamma_b^2-4\sqrt3\,gA\gamma_b=0 .
\label{eq:pole_discriminant}
\end{equation}
Because $g<0$, the positive low-threshold solution is
$A=(4-2\sqrt3)|g|\gamma_b$, hence
\begin{equation}
\phi_{b,1}^2(\kappa^\ast)
=(8-4\sqrt3)\,\omega_b\,
|g(\kappa^\ast)|\,\gamma_b(\kappa^\ast) .
\label{eq:beaks_threshold}
\end{equation}
Equation~\eqref{eq:beaks_threshold} compares the participation of the active
mode at the nonlinear coordinate with the damping-broadened background
through which that pole must bend the cusp contour.

\begin{figure*}[!tb]
\centering
\includegraphics[width=\textwidth]{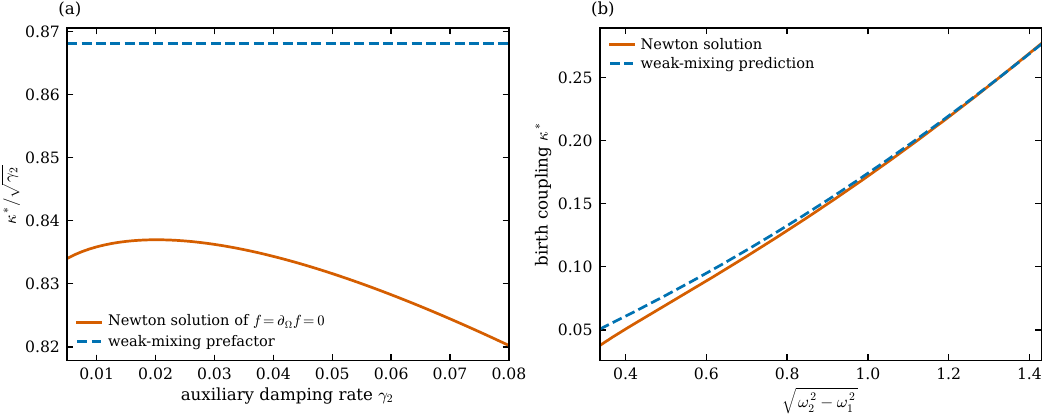}
\caption{Checks of the isolated-pole scaling in the modal two-pole model.
Solid curves are Newton solutions of Eq.~\eqref{eq:birthcondition}; dashed
curves are the weak-mixing asymptotics. (a)~Birth coupling divided by
$\sqrt{\gamma_2}$ over the damping sweep used in the text, against the
constant prefactor of Eq.~\eqref{eq:kstar_scaling}. (b)~Birth coupling versus
$\sqrt{\omega_2^2-\omega_1^2}$ at $\gamma_2=0.02$, compared with
Eq.~\eqref{eq:kstar_scaling} using $\omega_b\simeq\omega_2$ and
$\gamma_b\simeq\gamma_2$. The discrepancy grows as the weak-mixing condition
$2\kappa^\ast\ll\omega_2^2-\omega_1^2$ is lost.}
\label{fig:thresholdscaling}
\end{figure*}

For the baseline hardening upper-pole branch and weak modal mixing,
$2\kappa\ll\omega_2^2-\omega_1^2$,
$\phi_{b,1}^2\simeq\kappa^2/(\omega_2^2-\omega_1^2)^2$ and
$|g|\simeq1/(\omega_2^2-\omega_1^2)$. Equation~\eqref{eq:beaks_threshold}
then reduces to
\begin{equation}
\kappa^\ast\sim
\sqrt{(8-4\sqrt3)\,\omega_b\gamma_b
      (\omega_2^2-\omega_1^2)} .
\label{eq:kstar_scaling}
\end{equation}
Equation~\eqref{eq:kstar_scaling} is the full asymptotic statement; the
one-factor readings follow from it with the remaining factors held fixed.
Hence $\kappa^\ast\propto\sqrt{\gamma_b}$ at fixed mode separation, and
$\kappa^\ast\propto\sqrt{\omega_2^2-\omega_1^2}$ at fixed $\omega_b$ and
$\gamma_b$ in the weak-mixing regime. Newton solutions of Eq.~\eqref{eq:birthcondition} recover the damping
scaling to within about $5\%$ over $\gamma_2\in[0.005,0.080]$ and show the
expected crossover when weak mixing fails (Fig.~\ref{fig:thresholdscaling}).
Substituting the numerically evaluated participation and background into
Eq.~\eqref{eq:beaks_threshold} recovers the refined threshold to within a
few percent.

\subsection{Local catastrophe type}\label{ssec:recognition}
The projection-fold condition locates the birth but does not distinguish its
local topology. Put $u=\rho+cy$, with $c=3\beta_1/4$. Completing the square
in the fold equation gives $9(u-\rho/3)^2=f$. Combining this identity with
Eq.~\eqref{eq:beaksexp} yields the local critical set
\begin{equation}
9\,\delta u^2-a_\Omega\,\delta\Omega^2
=b_\kappa\,\delta\kappa .
\label{eq:beaksrecog}
\end{equation}
If $a_\Omega>0$, the quadratic form is indefinite: two fold branches touch and reconnect, leaving a gap bounded by the born cusps. If $a_\Omega<0$, the form is definite and an isolated
oval appears from an empty neighborhood, the lips-type geometry. The modal
two-mode birth has $a_\Omega=1370>0$ and is beaks type.

Written in receptance variables, this is the standard plane-to-plane
recognition. The projection of the fold surface has a corank-one singularity
whose discriminant has Hessian determinant proportional to $-a_\Omega$
\cite{Arnold1981,Kuznetsov2004,Saji2010}. This recognition connects the pole
threshold to the network-wide beaks/lips taxonomy of
Section~\ref{sec:cascade}.

\section{Full harmonic-balance validation}\label{sec:hbval}
The pole theory is derived from first-harmonic balance. We test it against
periodic solutions of the original equations, retaining the cubic-generated
harmonics and locating folds by pseudo-arclength continuation. Numerical
details, convergence tests, and the cusp-tip estimator are given in
Appendix~\ref{app:methods}. For the two-mode examples, $N_{\rm HARM}=3$ is
already converged at the cusp frequencies; the larger multimode tests use
$N_{\rm HARM}=5$ or $7$.

As a baseline numerical check, the ordinary lower cusp is predicted at
$\Omega_c=1.06731$ by the exact receptance and located at $1.06732$ by full
harmonic balance. Details of the tip estimator are given in
Appendix~\ref{app:folds}. The upper pair probes the approximation at larger
response amplitude: at $\kappa=0.20$, the modal receptance predicts
$1.31308$ and $1.36593$, while full harmonic balance gives $1.31248$ and
$1.36488$, errors of $0.046\%$ and $0.077\%$.

\subsection{Full-system birth and pair separation}\label{ssec:hbbeaks}
\begin{figure*}[!tbp]
  \centering
  \includegraphics[width=\textwidth]{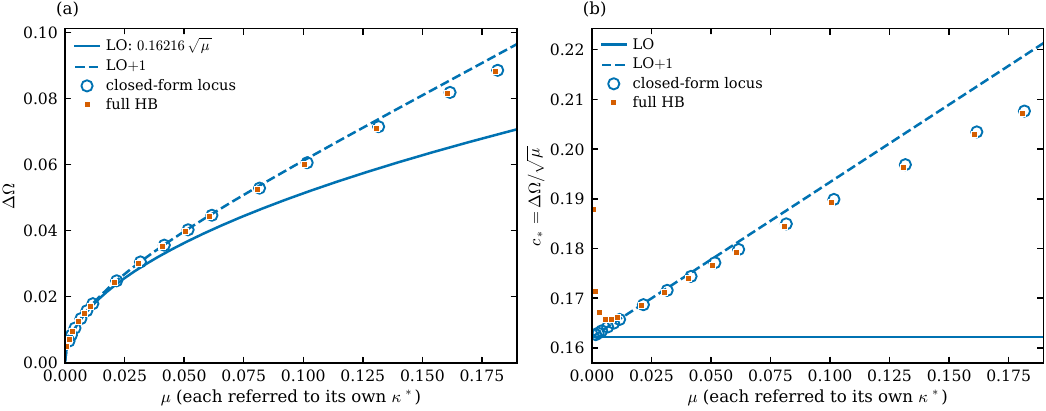}
  \caption{Square-root separation of the cusp pair at the beaks birth.
The full-HB data are referenced to the independently located
$\kappa^\ast_{\rm HB}=0.119322$; the reduced curve uses
$\kappa^\ast=0.1183654$. The dashed curve includes the next term in
Eq.~\eqref{eq:cstar_finite_mu}. The two closest points are the least
reliable because the cusp-tip fit windows overlap.}
  \label{fig:cuspofcusps}
\end{figure*}

The codimension-two birth is a stronger test. We locate it in the
full harmonic-balance problem without fitting the square-root law. Below the
birth, the bistable band near the active pole remains connected; above it,
an interval of frequency contains no bistability between two bistable
regions. Bisection on this topological change gives, after the two runs described below are
intersected,
\begin{equation}
\kappa^\ast_{\rm HB}\in[0.1193188,\,\,0.1193203],
\label{eq:kstar_hb}
\end{equation}%

compared with $0.1183654$ from the modal receptance and $0.11896$ from the
exact linear receptance. The remaining $0.30\%$ difference between the
exact-receptance prediction and the full-system birth is dominated by
effects beyond the first-harmonic organizer, not by the modal-damping
approximation. The convergence study of Appendix~\ref{app:projconv} returns $[0.1193188,\,0.1193250]$, and an independently configured single-setting run brackets the same point in $[0.1193096,\,0.1193203]$; both are reported in the Letter's Supplemental Material \cite{Sarkar2026Letter}. The test is one-sided: a detected gap bounds $\kappa^\ast$ from above firmly, whereas an undetected one bounds it from below only to the frequency-scan floor. The two runs are therefore intersected, not averaged, which gives \eqref{eq:kstar_hb}; the $0.30\%$ figure above is quoted against that interval. Figure~\ref{fig:cuspofcusps} and
Table~\ref{tab:beaks_hb} refer their data to the convergence-study midpoint
$0.119322$, which sits $1.7\times10^{-6}$ above the interval; over the couplings
plotted there this shifts $\mu$ by less than $0.1\%$ and does not affect the
quoted agreement.

The local expansion of $f$ also supplies the first finite-$\mu$ correction
to the square-root separation, with $\mu=\kappa-\kappa^\ast$,
\begin{equation}
c_\ast(\mu)=c_\ast^{(0)}+c_\ast^{(1)}\mu+O(\mu^2),
\label{eq:cstar_finite_mu}
\end{equation}
where $\Delta\Omega=c_\ast(\mu)\sqrt{\mu}$, with $c_\ast^{(0)}=0.16216$ and
$c_\ast^{(1)}=0.31199$. Appendix~\ref{app:finite_mu} gives the derivative formula for
$c_\ast^{(1)}$. Referred to the independently located
$\kappa^\ast_{\rm HB}$, the full-HB pair follows this prediction to
within $\pm1.2\%$ over $\kappa\in[0.125,0.18]$. Each curve is referred to
its own threshold, so this tests the predicted separation law and its
coefficient, not the absolute location of the birth, which
Eq.~\eqref{eq:kstar_hb} tests separately. The discrepancy reaches
$-5.2\%$ at $\kappa=0.30$, where the response is strongly anharmonic. The
two points closest to the birth are less reliable because the two cusp-tip
fit windows approach one another; they are displayed rather than absorbed
into a fitted threshold.

The complete cusp data are listed in Table~\ref{tab:beaks_hb}.

\section{Pole-resolved catastrophe ladder in multimode networks}\label{sec:cascade}
The two-mode birth suggests a network-level picture. For a real
$N$-degree-of-freedom second-order network, we call one complex-conjugate
modal pole pair a \emph{visible pole} when its residue in the driving-point
receptance is nonzero at the nonlinear coordinate. Let $P\le N$ denote the
number of such positive-frequency modal resonances. In the isolated-pole
regime, the primary visible pole carries the ordinary Duffing cusp and each
remaining visible pole is one local candidate
for an additional cusp-pair birth. This gives at most $P-1$ candidate rungs
when the poles are well separated and the background varies slowly across
each resonance. It is not a global pole-count theorem: a strongly
frequency-dependent complex background can change the number of stationary
points of the cusp contour.

\subsection{A resolved pole ladder}\label{ssec:cascade}
Chains and hubs make the mechanism visible without a complicated
background. In a four-resonator hub with satellite frequencies
$(1.25,1.5,1.75)$, the three non-primary poles generate births at
$\kappa^\ast\simeq0.096$, $0.167$, and $0.234$, so the cusp count rises
$1\to3\to5\to7$ (Fig.~\ref{fig:cascade}). Each onset lies near the
corresponding linear pole, and the higher poles require stronger coupling to
reach the physical cusp contour. Nearest-neighbor chains show the same
sequence at larger onset couplings. The ordering is consistent with the
threshold in Eq.~\eqref{eq:beaks_threshold}: a pole must acquire enough
participation at the nonlinear coordinate to overcome damping and the local
background.

\subsection{Random-network test of the ladder}
We next test the candidate count on $100$ random partially connected
networks with $N=3$--$6$ (Appendix~\ref{app:random}). On-site frequencies are drawn from
$[0.6,2.4]$, damping-rate seeds from $[0.01,0.04]$, and each possible coupling is
present with probability $0.7$ and a random weight in $[0.3,1.0]$. Seven
realizations contain a mode dark at the nonlinear coordinate, so that
$P<N$.

The relevant quantity is the largest number of additional pairs that
coexist at one parameter value, not the number of birth events encountered
during a sweep. Pair creation and annihilation make the cusp count
nonmonotonic in $60$ of the $100$ networks. With this distinction, no
realization exceeds $P-1$ coexisting additional pairs, and $23$ realize the
full ladder. The coarse sweep contains $236$ onset intervals. One interval
appears to create four cusps at once; refining that coupling interval by a
factor of $400$ resolves two distinct pair births at $\kappa=0.63912$ and
$0.64107$, associated with different poles. The resolved ensemble
contains $237$ birth events and $474$ born cusps, with every event creating
one pair. This numerical evidence supports the isolated-pole ladder but does
not replace the hypotheses behind it. Neither the observed ceiling nor the type-rule
accuracy of Section~\ref{ssec:typevalidity} is an artifact of the sampling:
an independent seed, a doubled coupling grid, a tightened cusp-association
tolerance, and a rebuild on the exact non-proportionally damped receptance each
leave no realization above $P-1$ and the rule accurate to $97.2$--$99.4\%$
(Appendix~\ref{app:random}). The ensemble is used below mainly to
determine when the pole-background type shortcut fails.

\subsection{Type rule and its range of validity}\label{ssec:typevalidity}
For a generic cusp-pair birth in this scalar first-harmonic class, the
local type is determined by the curvature $a_\Omega$ in
Eq.~\eqref{eq:beaksrecog}. For an isolated pole on a slowly varying,
nearly real background $G_a$, the local expansion further gives
\begin{equation}
\operatorname{sign}a_\Omega
\simeq-\operatorname{sign}\operatorname{Re}G_a(\omega_b).
\label{eq:typecrit}
\end{equation}
This follows from the pole expansion of Eq.~\eqref{eq:polebackground}. Let
$\tilde f\eqdef\operatorname{Re}G-\sqrt3\operatorname{Im}G$ be the factor of
$f=\rho^2-3\sigma^2$ that vanishes on the hardening contour, so that
$f=(\rho-\sqrt3\sigma)\,\tilde f/|G|^2$ identically. At a birth $\tilde f$
and $\partial_\Omega\tilde f$ vanish while the prefactor does not, so only the
term in which both derivatives act on $\tilde f$ survives, and with
$\rho-\sqrt3\sigma=2\rho$ on the contour,
\[
a_\Omega=\frac{\rho_\ast}{|G_\ast|^2}\,\partial_\Omega^2\tilde f ,
\]
where $\rho_\ast$ is $\rho$ at the birth, negative on the hardening branch by
$c\rho<0$. Multiplying
$\tilde f$ by $\delta^2+\gamma_b^2$ returns the left side of
Eq.~\eqref{eq:pole_quadratic} with $g$ replaced by
$\operatorname{Re}G_a-\sqrt3\operatorname{Im}G_a$. That quadratic and its
first derivative also vanish at the birth, so its second derivative gives
$\partial_\Omega^2\tilde f$ directly and
\begin{equation}
a_\Omega\simeq
\frac{2\rho_\ast\bigl(\operatorname{Re}G_a
       -\sqrt3\operatorname{Im}G_a\bigr)}
      {|G_\ast|^2\,(\delta_\ast^2+\gamma_b^2)} .
\label{eq:aomega_pole}
\end{equation}
Discarding $\operatorname{Im}G_a$ against $\operatorname{Re}G_a$ gives
Eq.~\eqref{eq:typecrit}. Two approximations enter and are separated below: the background is held constant across the pole, and its phase is neglected; only the sign of $a_\Omega$ is claimed. On the softening branch the conjugate
factor vanishes and $\rho_\ast>0$, so for either sign of the cubic
$\operatorname{sign}a_\Omega=-\operatorname{sign}\beta_1\,
\operatorname{sign}\operatorname{Re}G_a(\omega_b)$.

In the two-mode family this shortcut changes sign as the auxiliary pole
crosses the driven resonance: poles below it give lips-type births and poles
above it give beaks-type births (Table~\ref{tab:typecriterion}).

\begin{table}[H]
\centering
\caption{Exact curvature classifier as the auxiliary pole is moved through
the primary resonance of the two-mode absorber. Values are computed from the
exact physical-coordinate receptance, not the modal form of
Eq.~\eqref{eq:closed_form}; at $\omega_2=1.25$ the latter gives
$\kappa^\ast=0.1183654$ and $a_\Omega=1370$. In this hardening,
nearly-real-background family, an auxiliary pole below the primary resonance
gives \emph{lips} and one above it gives \emph{beaks}. Each tabulated
$\omega_2$ has exactly one birth; the family has three only inside
$\omega_2\in(1.0497,1.0628)$, the window opened by the fold of the birth
locus treated in Sec.~\ref{ssec:typeboundary}.}
\label{tab:typecriterion}
\begin{tabular}{cccc}
\toprule
$\omega_2$ & $\kappa^\ast$ & $a_\Omega$ & type \\
\midrule
0.60 & 0.4458 & $-282$   & lips \\
0.80 & 0.3926 & $-180$   & lips \\
0.90 & 0.3388 & $-123$   & lips \\
0.95 & 0.2998 & $-94.2$  & lips \\
\midrule
1.10 & 0.0623 & $+171$   & beaks \\
1.25 & 0.1190 & $+1315$  & beaks \\
1.40 & 0.1679 & $+3823$  & beaks \\
\bottomrule
\end{tabular}
\end{table}

The approximation in Eq.~\eqref{eq:typecrit} neglects the phase of the
background. To quantify that limitation, define
\begin{equation}
\psi=\frac{|\operatorname{Im}G_a|}{|\operatorname{Re}G_a|} .
\label{eq:psidef}
\end{equation}
Across the resolved ensemble, evaluating the shortcut at the individual
cusps gives $462$ agreements with the exact curvature out of $474$
evaluations. The two cusps associated with one birth are paired observations,
so these counts are used as a diagnostic stratification rather than as
independent statistical trials. All $417$ cusp evaluations with
$\psi<0.2$ agree with the exact curvature; the agreement rate falls once the
background phase becomes larger (Table~\ref{tab:typevalidity}). Stratifying by
$|\operatorname{Re}G_a|$ does not produce the same separation, so the
failures are associated with a genuinely complex background rather than
merely a small real part. The value $\psi\simeq0.2$ is a useful
empirical boundary for this ensemble, not a universal threshold.

Equation~\eqref{eq:aomega_pole} bounds how much of that failure the neglected
phase can account for: it reverses the sign only for $\psi>1/\sqrt3$ with
$\operatorname{Im}G_a$ and $\operatorname{Re}G_a$ of the same sign. Six
ensemble cusps satisfy $\psi>1/\sqrt3$; two of them are among the twelve
failures, and retaining the imaginary part, that is, using
$\operatorname{sign}(\operatorname{Re}G_a-\sqrt3\operatorname{Im}G_a)$,
repairs exactly those two and raises the count to $464/474$. The other ten
failures lie at $0.31\le\psi\le0.54$, below that bound, so they are not
explained by the background phase at fixed background; they belong to the
constant-background approximation, which degrades in the same regime.

These counts are computed on an ensemble whose receptance is built from real
modal residues, which is the regime the shortcut assumes. The unbiased test is
the non-proportionally damped rebuild of Appendix~\ref{app:random}, where the
background carries a complex residue by construction: there the rule holds at
$457/470$, and the born cusps shift toward larger $\psi$, with $46$ rather
than $37$ in $[0.2,0.4)$. The accuracy is not an artifact of the modal construction, although the modal ensemble is the more favorable of the two. When
the shortcut is doubtful, the full curvature of $f$ remains the classifier.

\begin{table}[H]
\centering
\caption{Diagnostic stratification of the isolated-pole
background-sign rule versus
$\psi=|\operatorname{Im}G_a|/|\operatorname{Re}G_a|$ for the $474$
individual cusp evaluations associated with the resolved births in $100$
random networks. The paired cusps from one birth are not independent
statistical trials. The $\psi<0.2$ boundary is empirical for this ensemble,
not a universal constant.}
\label{tab:typevalidity}
\begin{tabular}{cccc}
\toprule
$\psi = |\operatorname{Im}G_a|/|\operatorname{Re}G_a|$ & cusps & rule holds & accuracy \\
\midrule
$[0.00,\,0.05)$ & $136$ & $136$ & $100\%$ \\
$[0.05,\,0.10)$ & $171$ & $171$ & $100\%$ \\
$[0.10,\,0.20)$ & $110$ & $110$ & $100\%$ \\
\midrule
$[0.20,\,0.40)$ & $37$  & $33$  & $89\%$  \\
$[0.40,\,0.80)$ & $14$  & $8$   & $57\%$  \\
$[0.80,\,\infty)$ & $6$ & $4$   & $67\%$  \\
\bottomrule
\end{tabular}
\end{table}

\subsection{Full-equation realization of both local types}\label{ssec:cascadehb}
Multi-harmonic continuation confirms both reconnections in the original
equations. A three-resonator hub undergoes a beaks-type birth: above the
predicted threshold an interval with no bistability opens inside a previously
connected band. At $\kappa=0.14$, the full-HB gap is bounded within the
frequency-scan resolution by the cusp pair predicted from the exact linear
receptance. With the auxiliary mode below the drive ($\omega_2=0.95$), the
two-mode absorber realizes the lips case. The exact receptance predicts
$\kappa^\ast=0.2998$; below it no bistability is found in the local scan,
whereas above it an isolated lens appears. At $\kappa=0.333$ the resolved
lens spans $\Omega\in[1.01982,1.06454]$, compared with the predicted
$[1.01941,1.06495]$.

Representative Floquet calculations also confirm that the two outer
periodic branches are stable at the tested points. In the hub beaks their
spectral radii are about $0.91$--$0.92$; in the $\omega_2=0.95$ lips case
they are $0.963$ and $0.982$, with the intervening branch unstable. These
checks establish bistable coexistence at representative points, not a global
stability theorem for every point inside the fold regions.

\subsection{The classifier at its weakest point}\label{ssec:typeboundary}
Those two realizations sit where the classifier is safe: $|a_\Omega|$ is
$1315$ at the benchmark birth, so no plausible correction reaches the sign.
The informative test is the fold of the birth locus, where $a_\Omega$ passes
through zero and the two topologies coexist over a window narrower than the
resolution of any full-system scan. At $\omega_2=1.06$ the family carries two
births of opposite predicted type, a beaks at
$(\Omega^\ast,\kappa^\ast)=(1.0785,0.04100)$ with $a_\Omega=+40.0$ and a lips
at $(1.0611,0.05922)$ with $a_\Omega=-18.6$, curvatures $33$ and $71$ times
smaller than at the benchmark birth. Both sit at $\varepsilon\simeq0.1$ and
$\eta_3\simeq3\times10^{-3}$, well inside the regime where the
fundamental-harmonic reduction is quantitative, so a failure here is a failure
of the classifier and not of the reduction. That separation is what makes the
point worth testing.

\begin{figure*}[!tb]
\centering
\includegraphics[width=\textwidth]{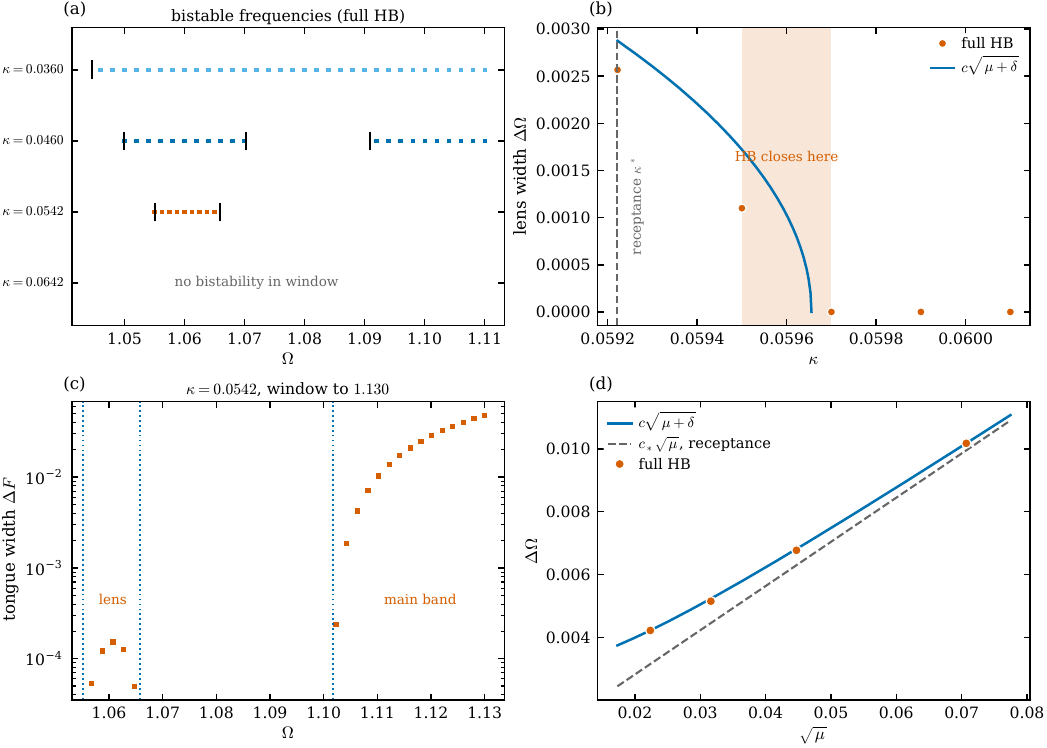}
\caption{The reconnection sequence and the beaks/lips classifier at the fold of
the birth locus, $\omega_2=1.06$. Blue curves and vertical marks are predicted
from the linear receptance before any nonlinear computation; orange points are
multi-harmonic balance of the original equations at $N_{\rm HARM}=5$.
(a)~Bistable frequencies along one coupling sweep: connected at
$\kappa=0.0360$, a gap straddling the predicted beaks frequency at $0.0460$, an
isolated lens at $0.0542$, none at $0.0642$. Band edges fall on the predicted
cusps to within the scan step. (b)~Lens width against coupling: the receptance
places the birth at $\kappa^\ast=0.0592217$ (dashed), where the lens is still
$2.6\times10^{-3}$ wide; it closes in $[0.0595,0.0597]$, low by about $0.7\%$.
(c)~On a window reaching $1.130$ the lens is separated from the main band, the
third predicted cusp lying on the upper band edge. (d)~Separation law:
$c\sqrt{\mu+\delta}$ with $c=0.1381$, $\delta=4.34\times10^{-4}$, against
$c_\ast=0.1408$ from the receptance; the displaced threshold, not a changed
exponent, accounts for the offset from the dashed line.}
\label{fig:typeboundary}
\end{figure*}

Figure~\ref{fig:typeboundary} follows the whole reconnection sequence through
that window in the original equations. Increasing the coupling at fixed
$\omega_2$ takes a connected bistable band to a band with a gap straddling the
predicted beaks frequency, then to an isolated lens, then to nothing; every
edge falls on a cusp predicted from the linear receptance alone, to within the
frequency-scan step. The lens is separated from the main band by a clear
interval on a window wide enough to contain the third physical cusp, so it is
isolated rather than a resolved end of a larger structure. Its width follows
$\Delta\Omega=c\sqrt{\mu+\delta}$ with $c=0.1381$ against the receptance value
$0.1408$, an agreement of $1.9\%$ at a curvature seventy times below the
benchmark birth.

The same data locate a limit the reduced description does not carry. The lens
survives at $\kappa^\ast$ itself, where the receptance predicts no cusps at
all, and closes only in $[0.0595,0.0597]$: the organizer places this birth
about $0.7\%$ low in $\kappa$, against $0.30\%$ at the benchmark birth. Location
degrades while the topology and its separation law hold, which is the behavior
Sec.~\ref{ssec:cascadelimits} anticipates, measured here at the curvature
where the classifier is least protected. Whether the third-harmonic correction
displaces the boundary is below the resolution of this test and is not claimed.
The same runs underlie a shorter check in the Letter's supplementary
material~\cite{Sarkar2026Letter}; what they establish here is that the
full-system reconnection can be read without leaving this paper.

\subsection{Where the leading-order location degrades}\label{ssec:cascadelimits}
The higher rungs tend to occur farther from the primary resonance and at
larger response amplitude. The first-harmonic geometry becomes
less quantitative even when the local topology remains recognizable. In the
lips family, moving the auxiliary frequency from $0.95$ to $0.60$ raises the
anharmonicity at birth from about $0.35$ to $0.69$. At $\omega_2=0.60$ the
full-HB onset lies roughly $15\%$ above the receptance prediction and the
predicted lens is only partly realized. By contrast, at $\omega_2=0.95$ the
birth threshold and lens edges remain accurate to a few tenths of a percent
although the response is already outside a strict weak-anharmonicity regime.
In these tests, the local topology persists beyond the regime in which
the leading-order location remains quantitatively accurate. Section~\ref{ssec:validity}
provides the corresponding harmonic-response diagnostic.

\section{Experimental signatures and platform realizations}\label{sec:fingerprint}
The normal forms suggest a direct experimental test. On a fixed-coupling
slice, a generic cusp closes as
$\Delta F_{\rm tongue}\propto|\Omega-\Omega_c|^{3/2}$. Across the
cusp-pair birth, the two cusp frequencies separate as
$\Delta\Omega\propto\sqrt{\kappa-\kappa^\ast}$. Either exponent can occur
in other settings, so their conjunction should be used only within a generic
two-control unfolding; a deliberately tuned higher-codimension singularity
can reproduce the same exponents. The beaks/lips subtype is read from the
reconnection itself: a gap opening inside an existing fold band is beaks
type, whereas an isolated lens born from an empty neighborhood is lips type.

In practice, the topology is easier to measure than
$a_\Omega=\tfrac12\partial_\Omega^2f$ directly. Estimating that second
derivative from a measured phase response requires smoothing and phase
unwrapping, while the reconnection can be identified from the presence or
absence of folds across frequency. Control-based continuation already allows
periodic responses to be followed through saddle-node folds and cusps
\cite{SieberEtAl2008,BartonSieber2013,NevilleEtAl2018}, and a cusp
catastrophe has been resolved directly in a silicon MEMS resonator
\cite{UkaZhao2026}. Detached states of the kind a lips-type birth creates
have likewise been observed, in a coupled-cavity platform with competing Kerr
nonlinearities, where they are reached by pulsed rather than swept excitation \cite{ZhangEtAl2026}. That is the same inaccessibility that makes an isolated lens easy to miss in a frequency sweep. A test of the present
birth therefore requires one additional control sweep: vary the network
parameter that moves the relevant pole, bracket the onset from below, then
locate the two cusp tips at several parameter values above it and test whether
their separation extrapolates to a common $\kappa^\ast$. The drive transfer function should also be checked for a zero
near the predicted frequency, because $G$ determines the intrinsic organizer
whereas $H$ determines whether the chosen port can reach it.

There is a separate experimental limit close to a cusp. Along a control
direction transverse to the cusp discriminant, the normal form gives a
cube-root response of the local amplitude to control detuning, so the
small-signal susceptibility grows as the tip is approached. At the same
time the activation barrier between coexisting states vanishes, making
noise-induced switching increasingly important \cite{Dykman2012}. The
practical resolution of a cusp is set by measurement time and
switching statistics as well as by the frequency step; this is one reason the
numerical estimators in Appendix~\ref{app:folds} approach the tip from
resolved folds rather than solving at the degeneracy itself.

\subsection{Nanomechanical scale}\label{ssec:nems}
Coupled nanomechanical resonators provide a natural mechanical setting:
Duffing bistability and tunable inter-resonator coupling are established in
closely related devices \cite{LifshitzCross2008,AldridgeCleland2005,
AlmogEtAl2007,KarabalinRoukes2009,EkinciRoukes2005}. A lightly damped mode
with quality factor $Q$ has $\gamma/\omega\simeq(2Q)^{-1}$, so the
$\sqrt{\gamma_b}$ scaling of Eq.~\eqref{eq:kstar_scaling} moves the
pole-induced birth to weaker coupling as $Q$ rises. A quantitative device
prediction should use the measured receptance, because the mapping of
coupling and Duffing coefficients into the nondimensional model is
platform-specific.

\subsection{Coupled-mode reduction and a superconducting design estimate}\label{ssec:cmt}
The rank-one reduction is algebraic and also applies to standard
steady-state coupled-mode equations. For linearly coupled modes in a frame
rotating at the drive, with a real Kerr coefficient $K$ on mode 1,
\begin{equation}
F\,\frac{H(\Omega)}{G(\Omega)}
 = A_1\left[\frac{1}{G(\Omega)}+K|A_1|^2\right],
\label{eq:cmt_reduced}
\end{equation}
with $G=e_1^T\mathcal Z^{-1}e_1$ and $H=e_1^T\mathcal Z^{-1}d$. Here
$\mathcal Z$ contains detunings, losses, and linear couplings, and $d$
selects the drive port. This is the same scalar equation as
Eq.~\eqref{eq:receptance_drive} under $3\beta_1/4\to K$. A complex Kerr
coefficient or a second nonlinear mode lies outside this model. The use of a
linear environment seen by a Josephson element is consistent with black-box
circuit descriptions \cite{NiggEtAl2012}; the driven Kerr response itself is
standard in circuit and cavity settings \cite{RobertsClerk2020}.

As a concrete scale, take a SQUID-terminated resonator with
$f_1=5.17$~GHz, linewidth $\kappa_1/2\pi=5.0$~MHz, and
$K/2\pi=-80$~kHz, coupled to an auxiliary mode with
$\kappa_2/2\pi=0.50$~MHz. The negative Kerr selects the softening contour,
$\arg G=-30^\circ$, so in this two-mode geometry the useful auxiliary pole
lies below the nonlinear resonance. Placing it $50$~MHz below gives
$J^\ast/2\pi=2.475$~MHz and $f_d-f_1=-50.07$~MHz. The cusp occupation is
$367$ photons. Post-birth separation is quoted at a single larger coupling:
at $J/2\pi=5$~MHz the two cusp frequencies are $0.99$~MHz apart, about twice
the auxiliary linewidth. Every other quantity in this subsection is evaluated
at the birth, $J=J^\ast$. One feature of that operating point drives the rest:
the drive lands $0.066$~MHz below the auxiliary resonance, so the auxiliary
denominator $D_2=\delta_2+i\kappa_2/2$, with $\delta_2$ the auxiliary detuning
from the drive, is only $|D_2|/2\pi=0.259$~MHz---half a linewidth. These numbers are an illustrative coupled-mode design, not
universal thresholds.

The fold geometry and dynamical stability must again be distinguished. At
this operating point the upper beaks sub-wedge contains a substantial region
with two stable outer branches, but the lower sub-wedge contains almost none
in the parameter scan considered. The receptance locates and
classifies the fold reconnection correctly even when only part of that fold
geometry is visible as stable steady-state bistability. This is why the main
theory is stated for the period-one fold set rather than for stability.

This asymmetry is not confined to the quoted operating point. In the
coupled-mode scan used for this realization,
$\kappa_2/2\pi\in[0.25,4]$~MHz and the auxiliary-mode detuning spans
$20$--$80$~MHz in magnitude; the upper sub-wedge remains partly usable
throughout, whereas the lower sub-wedge contains essentially no region with
two stable outer branches. This scan is specific to the illustrative device
model, but it shows that the stability asymmetry is not a single-point
artifact.

The drive port changes the required power but not the intrinsic organizer.
For the two-mode matrix
$\mathcal Z=\bigl(\begin{smallmatrix}A&-J\\-J&B\end{smallmatrix}\bigr)$,
a drive on the auxiliary mode gives $G=B/\det\mathcal Z$ and
$H=J/\det\mathcal Z$ with $B=D_2$, so $|H/G|=J/|D_2|$. At the birth this is
$9.6$. The required drive amplitude in the
normalization of Eq.~\eqref{eq:cmt_reduced} is lower by that factor. For
equal external port normalizations, this corresponds to a $19.6$~dB reduction
in input power relative to direct drive of the nonlinear mode. The pole enhancement also
raises the auxiliary occupation: at the birth,
$n_2/n_1=J^2/|D_2|^2\simeq92$, giving about $3.4\times10^4$ photons in the
nominally linear mode for $n_1=367$. The small $|D_2|$ is the origin of both
factors.

That large auxiliary population makes parasitic nonlinearity a relevant
constraint. With reciprocal cross-Kerr $\chi$, first-order elimination of
the auxiliary amplitude gives
\begin{equation}
K_{\rm eff}=K+\chi J^2
\left(\frac{1}{|D_2|^2}+\frac{1}{D_2^2}\right).
\label{eq:rankone_tol}
\end{equation}
At the birth, with $J=J^\ast$ and $D_2/2\pi=(0.066+0.25i)$~MHz, the
correction coefficient $J^2(|D_2|^{-2}+D_2^{-2})$ is $12.0-45.4i$, so the
largest effect is an amplitude-dependent loss rather than a conservative
Kerr shift. Requiring that induced loss to remain below $5\%$ of $|K|$
gives the design-specific tolerance $\chi/|K|\lesssim10^{-3}$. The same
pole that makes a cusp-pair birth easy to access therefore amplifies
nonlinearity carried by the auxiliary mode. A nominally linear auxiliary
resonator, or an independently characterized correction to
Eq.~\eqref{eq:cmt_reduced}, is required in this regime.

\section{Discussion}\label{sec:discussion}
The organizing object is the linear response presented to the coordinate
that carries the nonlinearity.
For a single localized cubic, first-harmonic balance reduces that environment
to the driving-point receptance $G$. The pole decomposition then resolves the
mechanism behind cusp-pair births: a visible pole must acquire sufficient participation to
bend the cusp contour through a projection fold, and in the isolated-pole
regime Eq.~\eqref{eq:beaks_threshold} gives the threshold explicitly in terms
of participation, damping half-width, and background receptance under the
isolated-pole assumptions. Repeating the argument pole by pole produces the
catastrophe ladder seen in the chains, hubs, and random networks.

Within first-harmonic balance, the scalar reduction, cusp condition,
projection-fold condition, and curvature classifier are exact. The
closed-form pole threshold further assumes a lightly damped modal pole on a
slowly varying, nearly real background. Its background-sign shortcut becomes
unreliable as that background acquires appreciable phase, as quantified by
the random-network ensemble. That ensemble conclusion is not an artifact of
its modal construction: rebuilding it on the exact non-proportionally damped
receptance moves the receptance at the born cusps by a median of $2.5\%$
(maximum $19\%$), and leaves the $P-1$ ceiling and the type-rule accuracy
unchanged (Appendix~\ref{app:random}). At the fold of the birth locus, where
the curvature is $33$ and $71$ times smaller than at the benchmark birth, the full
equations reproduce the entire reconnection sequence and the separation law to
$1.9\%$, while placing the birth itself about $0.7\%$ low in coupling
(Fig.~\ref{fig:typeboundary}). Multi-harmonic continuation exposes a different
limit: for strongly anharmonic higher rungs, the predicted location degrades
before the local beaks/lips geometry disappears. Stability is separate again;
the Josephson example retains the fold reconnection even where only part of
it survives as stable steady-state bistability.

\subsection{Scope and open directions}\label{ssec:open}
The scalar construction relies on one nonlinear direction. A second nonlinear
coordinate, a cross-cubic term, or nonlinear dissipation replaces the scalar
closure by a genuinely vector nonlinear balance. Overlapping poles also
invalidate the simplest pole--background formulas even though the exact
first-harmonic receptance criterion remains available. Extending the
closed-form threshold to such cases would require keeping the interacting
poles or nonlinear coordinates together rather than assigning a birth to one
isolated pole.

Linear coupling and mode placement control which poles are visible at the
nonlinear coordinate, their participation, and the background against which
they act. These quantities set the isolated-pole threshold before a nonlinear
continuation is performed. The drive port enters separately through $H$ and
changes the external forcing required to reach an intrinsic organizer.

Experimental tests need not reconstruct every derivative of the receptance.
A small-signal measurement can first identify the relevant pole and cusp
contour; continuation of the nonlinear response can then determine whether a
gap opens or an isolated lens appears as the network parameter crosses the
predicted onset. Control-based continuation already resolves nonlinear folds
and cusps in mechanical systems \cite{SieberEtAl2008,BartonSieber2013,
NevilleEtAl2018}. Similar questions arise in photonic and superconducting
resonator networks, where multistability is actively engineered
\cite{FitzpatrickEtAl2017,WengEtAl2020,RobertsClerk2020,SanyalEtAl2025,
MarisovaEtAl2025,ZhangEtAl2026}. The central restriction is structural rather
than platform-specific: one localized conservative cubic embedded in an
otherwise linear network.

A visible linear pole is a candidate organizer of an additional
cusp pair. Its participation and damping set the birth threshold; the
surrounding receptance sets the local type. In this restricted nonlinear
class, and at fundamental-harmonic order, the local linear receptance
organizes the fold catastrophes of the network.

\begin{acknowledgments}
I thank Professor Axel Gro\ss{} (Universit\"at Ulm) and Professor Pranab
Sarkar (Visva-Bharati University) for their guidance and for many valuable
discussions. ChatGPT (free version) and Anthropic Claude were used for
language editing, and Claude additionally for drafting analysis and
figure-rendering scripts; no figure image was generated or altered by a model.
All derivations and reported numbers were regenerated and checked by the author
from the deposited scripts. I acknowledge support from the
state of Baden-W\"urttemberg through bwHPC and from the German Research
Foundation (DFG) through grant INST 40/575-1 FUGG (bwForCluster JUSTUS~2), as
well as from the Dr.~Barbara Mez-Starck Foundation.
\end{acknowledgments}

\appendix

\section{Finite-distance expansion of the cusp-pair separation}
\label{app:finite_mu}
The next coefficient in Eq.~\eqref{eq:cstar_finite_mu} follows from the
Taylor series of the first-harmonic cusp condition. Write
$\delta=\Omega-\Omega^\ast$, $\mu=\kappa-\kappa^\ast$, and expand
\begin{equation}
f(\delta,\mu)=
a\delta^2+b\mu+c\delta^3+d\delta\mu
+e\delta^4+h\delta^2\mu+j\mu^2+\cdots ,
\label{eq:f_taylor_finite_mu}
\end{equation}
where all derivatives are evaluated at the projection fold and
\[
a=\tfrac12 f_{\Omega\Omega},\quad b=f_\kappa,\quad
c=\tfrac16 f_{\Omega\Omega\Omega},\quad d=f_{\Omega\kappa},
\]
\[
e=\tfrac1{24}f_{\Omega\Omega\Omega\Omega},\quad
h=\tfrac12 f_{\Omega\Omega\kappa},\quad
j=\tfrac12 f_{\kappa\kappa}.
\]
The two roots have the form
\begin{equation}
\delta_\pm=
\pm s_0\mu^{1/2}+s_1\mu\pm s_2\mu^{3/2}+O(\mu^2),
\end{equation}
with
\begin{align}
s_0^2&=-\frac{b}{a},\\
s_1&=-\frac{c s_0^2+d}{2a},\\
s_2&=-\frac{
a s_1^2+3c s_0^2s_1+d s_1+e s_0^4+h s_0^2+j
}{2a s_0}.
\end{align}
Hence
\begin{equation}
\Delta\Omega
=2s_0\sqrt{\mu}+2s_2\mu^{3/2}+O(\mu^{5/2}),
\end{equation}
so that $c_\ast^{(0)}=2s_0$ and $c_\ast^{(1)}=2s_2$.
Evaluating these derivatives on the modal two-mode cusp condition gives
$c_\ast^{(0)}=0.16216$ and $c_\ast^{(1)}=0.31199$, the values used in
Section~\ref{ssec:hbbeaks}.

\begin{table}[H]
\centering
\caption{Upper-pair cusp tips from full harmonic balance at
$\beta_1=0.147$. The full-system threshold
$\kappa^\ast_{\rm HB}=0.119322$ is obtained independently by topological
bisection. $c_\ast^{\rm HB}=\Delta\Omega/\sqrt{\kappa-\kappa^\ast_{\rm HB}}$ is compared with the analytic
leading-plus-next-order prediction.} 
\label{tab:beaks_hb}
\begin{tabular}{cccccc}
\toprule
$\kappa$ & $\mu$ & $\Omega_-^{\rm HB}$ & $\Omega_+^{\rm HB}$ &
$c_\ast^{\rm HB}$ & $c_\ast^{\rm LO+1}$ \\
\midrule
0.1200 & 0.000678 & 1.29902 & 1.30391 & 0.1879 & 0.1624 \\
0.1210 & 0.001678 & 1.29838 & 1.30540 & 0.1714 & 0.1627 \\
0.1225 & 0.003178 & 1.29785 & 1.30727 & 0.1671 & 0.1632 \\
0.1250 & 0.005678 & 1.29743 & 1.30993 & 0.1658 & 0.1639 \\
0.1275 & 0.008178 & 1.29732 & 1.31231 & 0.1658 & 0.1647 \\
0.1300 & 0.010678 & 1.29737 & 1.31454 & 0.1662 & 0.1655 \\
0.1400 & 0.020678 & 1.29842 & 1.32265 & 0.1685 & 0.1686 \\
0.1500 & 0.030678 & 1.30016 & 1.33014 & 0.1712 & 0.1717 \\
0.1600 & 0.040678 & 1.30227 & 1.33734 & 0.1739 & 0.1749 \\
0.1700 & 0.050678 & 1.30461 & 1.34437 & 0.1766 & 0.1780 \\
0.1800 & 0.060678 & 1.30712 & 1.35128 & 0.1793 & 0.1811 \\
0.2000 & 0.080678 & 1.31248 & 1.36488 & 0.1845 & 0.1873 \\
0.2200 & 0.100678 & 1.31826 & 1.37831 & 0.1893 & 0.1936 \\
0.2500 & 0.130678 & 1.32728 & 1.39825 & 0.1963 & 0.2029 \\
0.2800 & 0.160678 & 1.33666 & 1.41802 & 0.2030 & 0.2123 \\
0.3000 & 0.180678 & 1.34307 & 1.43113 & 0.2072 & 0.2185 \\
\bottomrule
\end{tabular}
\end{table}

\section{Numerical methods}\label{app:methods}
We summarize the periodic-orbit, continuation, and stability computations so
that the results are reproducible from the equations of motion alone. The
linear part is carried by $Z(\Omega) = K - \Omega^2 M + i\Omega C$ (with
$M = I$ here), one coordinate $e_1$ driven and carrying the cubic, so the
time-domain equations are $M\ddot{x} + C\dot{x} + Kx + \beta_1 x_1^3 e_1 =
F\cos(\Omega t)\,e_1$; the two-mode absorber
\eqref{eq:eom1}--\eqref{eq:eom2} and the multimode chains and hubs of
Section~\ref{sec:cascade} are instances.

\subsection{Harmonic balance}\label{app:hbmethod}
We use an alternating frequency--time (AFT) residual. Each coordinate is
expanded as a truncated Fourier series,
\begin{equation}
x_j(t) = \sum_{m=1}^{N_{\rm HARM}}
   \bigl[a_{jm}\cos(m\Omega t) + b_{jm}\sin(m\Omega t)\bigr],
\label{eq:fourier}
\end{equation}
so the unknown is the coefficient vector $z = \{a_{jm}, b_{jm}\}$ of
dimension $2 N N_{\rm HARM}$. Sampling $x(t)$, $\dot{x}(t)$, $\ddot{x}(t)$
from \eqref{eq:fourier} on a uniform grid of $n_t$ points over one period,
evaluating the physical residual $r(t) = M\ddot{x} + C\dot{x} + Kx +
\beta_1 x_1^3 e_1 - F\cos(\Omega t)e_1$, and projecting it onto the
harmonics gives the algebraic residual $R(z;\Omega,F) = \int_0^{2\pi/\Omega}
r(t)\{\cos,\sin\}(m\Omega t)\,dt$, whose zeros are the periodic orbits. The
Jacobian $\partial R/\partial z$ is obtained by forward-mode automatic
differentiation of this map. We solve $R = 0$ to residual norm below
$10^{-10}$ with $n_t = 256$--$512$; the odd parity of
the pure cubic sends even harmonics to the solver floor, and the leading
cubic-generated line $3\Omega$ first enters the basis at $N_{\rm HARM} = 3$.
The two-mode results use $N_{\rm HARM} = 3$, converged to five digits at their
anharmonicity (Appendix~\ref{app:nharm}). The multimode networks of
Section~\ref{ssec:cascadehb} sit at the larger anharmonicity $\varepsilon$ of
their higher rungs, where the higher cubic-generated lines carry more
weight, so we retain $N_{\rm HARM} = 5$ for the three-resonator hub and
$N_{\rm HARM} = 7$ for the two-degree-of-freedom lips. Increasing the
truncation further leaves the reported fold forcings of
Section~\ref{ssec:cascadehb} unchanged to five digits (hub
$N_{\rm HARM} = 5 \to 9$, lips $7 \to 11$).

\subsection{Locating cusp tips}\label{app:folds}
Saddle-nodes away from a tip are located at fixed $\Omega$ from
\begin{equation}
R(z;F)=0,\qquad J(z;F)v=0,\qquad \ell^Tv=1,
\label{eq:hb_sn_augmented}
\end{equation}
where $J=\partial R/\partial z$ and $\ell$ fixes the null-vector
normalization. Newton solves this system in $(z,F,v)$. A cusp tip is a
degenerate saddle-node, so the augmented Jacobian becomes ill-conditioned
as the tip is approached.

Bracketing the tip by the frequency at which Newton on
\eqref{eq:hb_sn_augmented} ceases to converge requires no extra machinery
and is accurate to the width to which that boundary can be bisected. It is
nonetheless the wrong estimator: the boundary is set by the conditioning of
the augmented Jacobian rather than by the geometry of the fold surface, and
it therefore moves with the harmonic truncation. On the upper pair it
shifts by up to $5\times10^{-3}$ in $\Omega$ between $N_{\rm HARM} = 2$ and
$3$, four orders of magnitude beyond the bisection width, and not
monotonically. We therefore measure away from the tip and extrapolate.

At each $\Omega$ inside a bistable wedge the periodic-orbit branch is
traced by pseudo-arclength continuation in $F$
(Appendix~\ref{app:continuation}), which passes smoothly through both
turning points, and the two fold forcings are read from the sign changes of
the tangent's $F$-component. The tongue width $\Delta F_{\rm
tongue}(\Omega)$ obtained this way is converged in the arclength step to
five digits. Because the cusp normal form makes $(\Delta F_{\rm
tongue})^{2/3}$ linear in $\Omega$, a linear fit of $(\Delta F_{\rm
tongue})^{2/3}$ against $\Omega$ locates the tip by its intercept, using
only well-conditioned solves.

Two checks fix the systematics. The fit window is shrunk in stages and the
intercept accepted once successive windows agree; the residual spread is
reported as the tip uncertainty, and over the $32$ tips of
Table~\ref{tab:beaks_hb} it has median $1.2\times10^{-4}$ and worst case
$7.0\times10^{-4}$ in $\Omega$. The truncation is
varied independently: the tips are unchanged from $N_{\rm HARM} = 3$ to $5$
to better than $10^{-6}$, so $N_{\rm HARM} = 3$ is converged for this
measurement. As an absolute check, on the lower cusp, where the modal closed form gives
$\Omega_c = 1.06724$ and the exact receptance gives $1.06731$, the method
returns $1.06732$, a relative agreement of $7\times10^{-5}$, with $r^2 = 0.999995$ on the
$(\Delta F_{\rm tongue})^{2/3}$ fit and agreement between $N_{\rm HARM} =
3$ and $5$ to nine digits.

The codimension-two cusp-pair point of the reduced model is the solution of
$f = 0$, $\partial_\Omega f = 0$ on the closed-form receptance
(Section~\ref{ssec:beaksdef}), refined by two-dimensional Newton
(the Newton solve).

The same point is located in the \emph{full} system without imposing the
receptance birth condition or any separation exponent. The reduced model is
used only to choose a frequency search window. At each coupling we scan
$\Omega$ across that window, evaluate the fold count by the arclength
continuation above, and ask whether there exists an interior run of
frequencies carrying fewer than two folds, bounded on both sides by
bistable frequencies. Below the birth the band is connected and no such run
exists; above it, the gap between the two sub-wedges provides one. Bisecting
the coupling on this binary test brackets the projection fold to
$6.2\times10^{-6}$ in $\kappa$, at a cost of one $\Omega$-scan per bisection
step. The scan window is set from the closed-form pair where one exists and
from a fixed interval about the reduced-model birth frequency where it does
not, so the test remains defined on both sides of the reduced-model
threshold. Its resolution floor is set by the $\Omega$ sampling: a gap is
detected once it exceeds one grid step, which for the grids used here
corresponds to a bias below $10^{-6}$ in $\kappa$, far under the quoted
bracket.

\subsection{Continuation of bistable windows}\label{app:continuation}
The bistable regions of Section~\ref{ssec:cascadehb} are traced by pseudo-arclength
continuation in $F$ at fixed $\Omega$: from a converged orbit, a predictor
along the tangent (the null vector of $[\partial R/\partial z\,|\,\partial
R/\partial F]$) is corrected by Newton on $R = 0$ together with an arclength
constraint, which passes smoothly through the turning points that $F$- or
$\Omega$-stepping falls off. Folds are the sign changes of the tangent's
$F$-component; a slice carrying two or more folds is bistable, and its fold
forcings are the boundaries plotted.

\subsection{Floquet stability}\label{app:floquet}
The stability of an orbit follows from its monodromy. Reconstructing $x(t)$
from the harmonic-balance coefficients and integrating the variational
equation $\dot{\Phi} = D(t)\,\Phi$, $\Phi(0) = I$, over one period, with
\begin{equation}
D(t) = \begin{pmatrix} 0 & I \\
  -\bigl(K + 3\beta_1 x_1(t)^2\, e_1 e_1^{T}\bigr) & -C \end{pmatrix},
\label{eq:variational}
\end{equation}
gives the monodromy $\Phi(2\pi/\Omega)$; its spectral radius below unity
marks a dynamically stable orbit. This is the calculation reported in Section~\ref{ssec:cascadehb}.

\subsection{Random-network ensemble}\label{app:random}
For the $100$-network ensemble of Section~\ref{sec:cascade}, on-site
frequencies are drawn uniformly from $[0.6,2.4]$ and sorted with the nonlinear
coordinate normalized to unit frequency. Damping-rate seeds are uniform on
$[0.01,0.04]$ and are used to construct the modal damping rates below; the
term physical-coordinate damping is reserved for the exact non-proportional
rebuild. Each pair of coordinates is connected independently with
probability $0.7$; a present edge receives a weight uniform on $[0.3,1.0]$.
The cusp locus is traced as the common coupling scale is increased. At each
coupling, roots of the physical cusp condition are associated with roots on
the preceding step by frequency proximity; apparent multiple births are
recomputed on a locally refined coupling grid. The sweep uses $300$ coupling
values uniformly spaced on $\kappa\in[0,1.5]$, drawn with random seed
$20260811$; roots on adjacent steps are treated as the same cusp when their
frequencies differ by less than $5\times10^{-3}$, and the single interval that
appears to create four cusps at once is subdivided $400$-fold.

The pole decomposition is modal and is stated here in full, since the
quantities $P$, $G_a$ and $\psi$ of Section~\ref{sec:cascade} are read from
it. Diagonalizing the stiffness matrix at a given coupling as
$K=V\Lambda V^{T}$ gives modal frequencies $\omega_m=\Lambda_{mm}^{1/2}$,
modal damping rates $z_m=V_{:m}^{T}\,\mathrm{diag}(\zeta)\,V_{:m}$, and
participations $\phi_{m,1}^2=V_{1m}^2$, so that
\begin{equation}
G(\Omega)=\sum_m\frac{\phi_{m,1}^2}
   {\omega_m^2-\Omega^2+2iz_m\Omega} .
\label{eq:modalG}
\end{equation}
The ensemble network is defined by these modal data, so that
Eq.~\eqref{eq:modalG} is its exact receptance; the drawn per-coordinate rates
$\zeta$ generate the modal damping rates rather than acting as a
physical-coordinate damping matrix. For a network damped by
$\mathrm{diag}(\zeta)$ in the physical coordinates,
$V^{T}\mathrm{diag}(\zeta)V$ carries off-diagonal entries and
Eq.~\eqref{eq:modalG} would instead be a modal-damping approximation. A mode
is counted as visible when $\phi_{m,1}^2>10^{-10}$, a property of $K$ alone,
taken as the largest count over the coupling grid because a mode dark at
$\kappa=0$ can acquire participation as the coupling is turned on; $P$ is that
count. At a
born cusp of frequency $\Omega$, the active pole is the mode nearest in
frequency, $a=\arg\min_m|\Omega-\omega_m|$, and the background is the same
sum with that term removed,
$G_a=\sum_{m\neq a}\phi_{m,1}^2/(\omega_m^2-\Omega^2+2iz_m\Omega)$,
evaluated at the cusp frequency rather than at $\omega_a$; the two differ by
the detuning $\delta_\ast$, which is of order $\gamma_b$ at a birth. This is
the $G_a$ entering $\psi$ and Table~\ref{tab:typevalidity}.

The three numerical choices in this construction were varied one at a time.
An independent draw (seed $20260908$) gives $492$ born cusps over $243$ onset
intervals, no network above $P-1$ coexisting pairs, and the type rule holding
at $489/492$. Doubling the coupling grid to $600$ values gives $472$ cusps,
again no network above that ceiling, and $464/472$. These interval counts are not
refined as the baseline was, so a few intervals still contain more than one
birth and the cusp count exceeds twice the interval count. Tightening the association
tolerance from $5\times10^{-3}$ to $2\times10^{-3}$ reproduces the baseline
exactly. The event and cusp tallies move with the grid, as they must, while
the $P-1$ ceiling and the accuracy of the type rule do not.

The modal-damping construction was tested the same way. Repeating the whole
ensemble on the exact non-proportionally damped receptance
$G=e_1^{T}(K-\Omega^2I+2i\Omega\,\mathrm{diag}(\zeta))^{-1}e_1$, with the
background obtained by removing the active conjugate pole pair from the exact
pole--residue expansion of the corresponding quadratic eigenvalue problem,
changes the receptance at the born cusps by a median of $2.5\%$
($90$th percentile $7\%$, maximum $19\%$). That is not a negligible
perturbation, and the conclusions are nevertheless unchanged: $470$ born cusps
over $233$ unrefined onset intervals, again no realization above the $P-1$ ceiling of coexisting
pairs, the same $23$ full ladders, and the type rule holding at $457/470$ with
every one of the $404$ cusps at $\psi<0.2$ correct. All resolved birth
records are stored in the archived reproduction package.

\section{Supplementary checks}\label{app:checks}
This appendix collects three numerical checks supporting the main text:
$N_{\rm HARM}$ convergence of the full HB solver, convergence of the
located projection fold, and the damping/mode-separation scalings of the
isolated-pole threshold.

\subsection{$N_{\rm HARM}$ convergence}\label{app:nharm}
The leading cubic-generated line enters at $N_{\rm HARM}=3$. Direct
truncation tests on ordinary folds change amplitudes by about
$5\times10^{-3}$ when going from $N_{\rm HARM}=2$ to $3$ and by less than
the reported digits thereafter. The more relevant check is the location of
the upper-pair cusp tips.

Table~\ref{tab:nharmtips} checks the cusp-tip truncation directly, using the same
window-refined estimator as Table~\ref{tab:beaks_hb}. The tips move in the
fourth decimal between $N_{\rm HARM} = 2$ and $3$, by about
$2\times10^{-4}$ in $\Omega$, and are then stable to better than $10^{-6}$
through $N_{\rm HARM} = 5$, at each of three couplings spanning the
measured range. The pair separation inherits the same behavior, changing
by $0.5$--$1.8\%$ from $N_{\rm HARM} = 2$ to $3$ and by at most $10^{-6}$
thereafter.

\begin{table}[!htbp]
\centering
\caption{Truncation dependence of the upper-pair cusp tips at
  $\beta_1 = 0.147$, computed with the same window-refined
  $(\Delta F_{\rm tongue})^{2/3}$ estimator used for
  Table~\ref{tab:beaks_hb} (Appendix~\ref{app:folds}), so that the
  $N_{\rm HARM} = 3$ rows reproduce the corresponding entries of that table
  exactly. Values are converged from $N_{\rm HARM} = 3$ onward, the residual
  drift to $N_{\rm HARM} = 5$ being below $10^{-6}$ in $\Omega$.}
\label{tab:nharmtips}
\begin{tabular}{ccccc}
\toprule
$\kappa$ & $N_{\rm HARM}$ & $\Omega_-^{\rm HB}$ & $\Omega_+^{\rm HB}$ &
$\Delta\Omega$ \\
\midrule
0.13 & 2 & 1.297177 & 1.314662 & 0.017485 \\
0.13 & 3 & 1.297366 & 1.314535 & 0.017169 \\
0.13 & 4 & 1.297366 & 1.314535 & 0.017169 \\
0.13 & 5 & 1.297366 & 1.314535 & 0.017169 \\
\midrule
0.15 & 2 & 1.299982 & 1.330235 & 0.030253 \\
0.15 & 3 & 1.300162 & 1.330144 & 0.029982 \\
0.15 & 4 & 1.300162 & 1.330144 & 0.029982 \\
0.15 & 5 & 1.300162 & 1.330144 & 0.029982 \\
\midrule
0.20 & 2 & 1.312277 & 1.364956 & 0.052679 \\
0.20 & 3 & 1.312478 & 1.364882 & 0.052403 \\
0.20 & 4 & 1.312478 & 1.364882 & 0.052403 \\
0.20 & 5 & 1.312479 & 1.364882 & 0.052402 \\
\bottomrule
\end{tabular}
\end{table}

\subsection{Convergence of the located projection fold}\label{app:projconv}
The full-system birth is detected from a frequency gap on a discrete grid, so
we repeated the bisection while varying the frequency sampling, scan window,
arclength step, and harmonic truncation. Using $40$, $60$, or $90$ frequency
samples; windows equal to $0.25$, $0.35$, or $0.50$ of the reduced-model pair
separation; arclength steps $0.01$ or $0.005$; and
$N_{\rm HARM}=3$ or $4$ returned the same bracket,
$\kappa^\ast_{\rm HB}=0.119322$ with width $6.2\times10^{-6}$. The quoted
uncertainty is the resolution of the topological bisection across
all settings tested.

\subsection{Scaling checks for the isolated-pole threshold}
\label{app:thresholdscaling}
The scaling statements following Eq.~\eqref{eq:kstar_scaling} were checked
directly on the modal two-pole model, solving the cusp-fold conditions
$f=0$ and $\partial_\Omega f=0$ by continuation rather than fitting an
exponent. At fixed $(\omega_1,\omega_2,\gamma_1)=(1,1.25,0.015)$, varying
$\gamma_2$ from $0.005$ to $0.080$ moves the birth coupling from
$0.05897$ to $0.23199$. Over that interval
$\kappa^\ast/\sqrt{\gamma_2}$ varies from about $0.837$ to $0.820$, compared
with the weak-mixing prefactor
$\sqrt{(8-4\sqrt3)\,\omega_2(\omega_2^2-\omega_1^2)}=0.868$.

A second continuation varies the auxiliary frequency at
$\gamma_2=0.02$. The agreement with
$\kappa^\ast_{\rm WM}
=\sqrt{(8-4\sqrt3)\,\omega_2\gamma_2(\omega_2^2-\omega_1^2)}$
improves systematically as
$2\kappa^\ast/(\omega_2^2-\omega_1^2)$ decreases: it is within about $3\%$
once that ratio is below $0.4$ and within $0.4\%$ below $0.3$.
These checks test the asymptotic reduction of the first-harmonic cusp-fold
condition; Section~\ref{sec:hbval} separately tests that organizer against
the original equations.

\section*{Code and data availability}
The public reproduction package contains the harmonic-balance and
continuation solvers, cusp-locus and pole-threshold calculations,
random-network ensemble, Floquet analysis, device checks, and the data used
in the figures and tables. It is openly available in Zenodo at
\href{https://doi.org/10.5281/zenodo.22672005}{doi:10.5281/zenodo.22672005}. The
archive is shared with the companion Letter~\cite{Sarkar2026Letter}, the two
resting on one code base; its script-to-claim map (\texttt{reproduce.md})
labels each entry by the manuscript it supports.

\bibliographystyle{apsrev4-2}
\bibliography{refs_prx}

\end{document}